# Advanced Manufacturing with Renewable and Bio-based Materials: AI/ML workflows and Process Optimization

Rigoberto Advincula [1,2]* and Jihua Chen[1]

[1]Center for Nanophase Materials Sciences, Oak Ridge National Laboratory (ORNL),
1 Bethel Valley Road, Oak Ridge, TN 37830 USA

[2]Chemical Engineering, University of Tennessee at Knoxville
1512 Middle Dr, Knoxville, TN 37996

*Correspondent author: radvincu@utk.edu

## Abstract

Advanced manufacturing with new bio-derived materials can be achieved faster and more economically with first-principle-based artificial intelligence and machine learning (AI/ML)-derived models and process optimization. Not only is this motivated by increased industry profitability, but it can also be optimized to reduce waste generation, energy consumption, and gas emissions through additive manufacturing (AM) and AI/ML-directed self-driving laboratory (SDL) process optimization. From this perspective, the benefits of using 3D printing technology to manufacture durable, sustainable materials will enable high-value reuse and promote a better circular economy. Using AI/ML workflows at different levels, it is possible to optimize the synthesis and adaptation of new bio-derived materials with self-correcting 3D printing methods and in-situ characterization. Working with training data and hypotheses derived from Large Language Models (LLMs) and algorithms, including ML-optimized simulation, it is possible to demonstrate more field convergence. The combination of SDL and AI/ML workflows can be the norm for improved use of biobased and renewable materials towards advanced manufacturing. This should result in a faster and better correlation between structure, composition, processing, and properties (SCPP). More agentic AI tasks, as well as supervised and unsupervised learning, can be incorporated to continuously improve optimization protocols. Deep Learning (DL), Reinforcement Learning (RL), and Deep Reinforcement Learning (DRL) with Deep Neural Networks (DNNs) can be applied to more generative AI directions in both AM and SDL with bio-based materials.

## 1. Introduction: A General Overview of Manufacturing and Polymer Materials

Polymer materials are ubiquitously used in many manufacturing methods, primarily formative manufacturing (FM; molding, thermoforming, pultrusion, blown-film molding, etc.).[1] Commodity polymers used in consumer products, automotive, aerospace, and packaging are the most dominant polymer class, most commonly polyolefins and other step-growth polymerized plastics like polyesters. Polymeric materials and composites, compared to other materials, offer advantages such as lower cost, design flexibility, lighter weight, and high-volume production. However, a common complaint is the lack of recyclability or circularity as a materials class.[2] Upcycling has been the goal for the last few years in both mechanical and chemical recycling.[3,4] However, a few commercial breakthroughs and many environmental issues will prevent new synthetic and fossil-fuel-based polymers from entering manufacturing.[1]

The main goal of this article is to chart a course for the growing demand for sustainable and renewable polymer materials in manufacturing, leveraging advances in discovery science and advanced manufacturing tools[5]. Environmental concerns are a primary driver, and overcoming the limitations of petroleum-based polymers is an opportunity to advance a circular economy.[6–9] The biodegradable properties of natural polymers and fibers can be a weakness or taken as a challenge.[10,11] It is essential to overcome these limitations, especially with new materials and manufacturing advances.[12] Artificial intelligence and machine learning (AI/ML) will accelerate research and development of new materials and advances in manufacturing.[6–9] Figure 1 illustrates the architecture of a possible Digital Twin (DT) that enables AI to interact with and guide manufacturing activities in the real world.

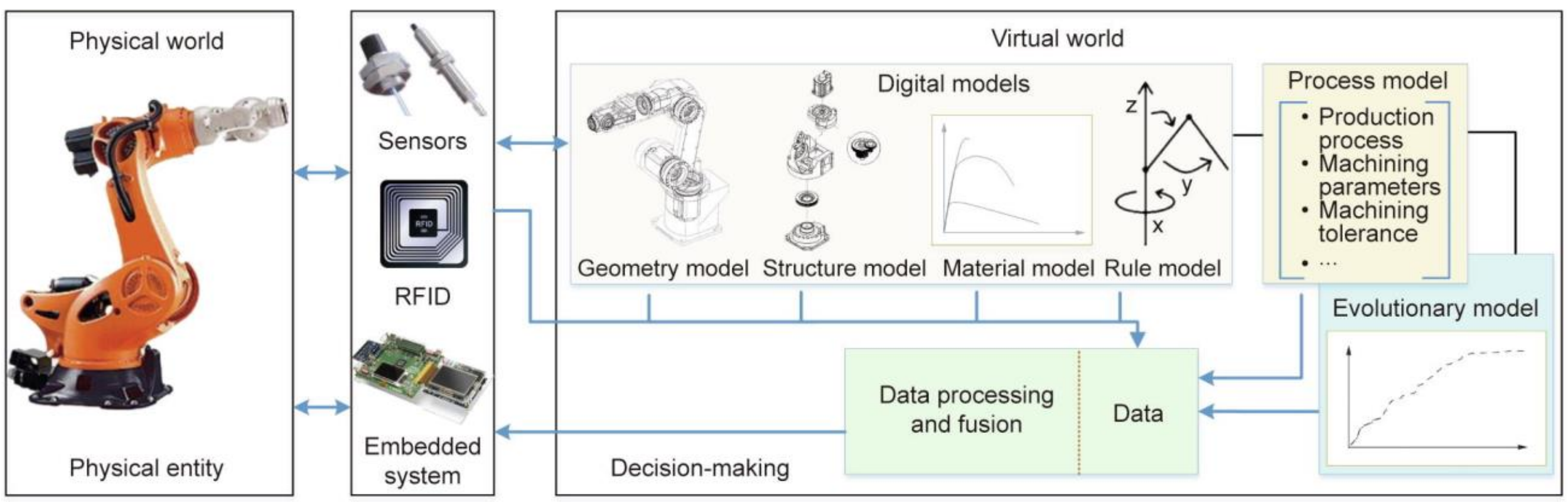

Figure 1. Models and data can be considered as the core elements of a DT. (Figures and Captions above are from reference,[13] without change, under Creative Commons BY NC ND 4.0 License.[14]) DT refers to digital twins.

## 2.Additive Manufacturing.

Additive manufacturing (AM) is a mechatronics-driven process that builds 3D objects layer by layer from a computer-aided design (CAD) model.[15] It is also commonly called 3D printing. Based on several approaches using raw powders, melt filaments, viscoelastic materials, photo-curable resins, etc., each will have different form-factor limits and functions that essentially control the mode of layer fabrication. More recent reviews are available to summarize this advanced manufacturing method for polymers.[16–18] While many reviews cover metals, ceramics, refractory materials, and other classes, this article focuses on polymers and bio-based renewable materials. A variety of thermoplastics, nanocomposites, fibers, nanofibers, thermosets, and elastomers have been used to improve processing and properties. Viscoelastic gel formulations and stimuli-responsive polymers are in demand from industry looking for new processing methods. Demands across many industrial processes warrant this development.[19]

Some popular 3D printing techniques are illustrated in Figure 2. These include fused deposition modeling (FDM), stereolithography (SLA), digital light projection (DLP), selective laser sintering ( SLS), and polymer jetting (Polyjet). Direct ink writing (DIW) is not in Figure 2. It is similar to FDM, but uses a gel or liquid-like ink instead of filament. Table 1 compares these 3D printing technologies in terms of their strengths and limitations.

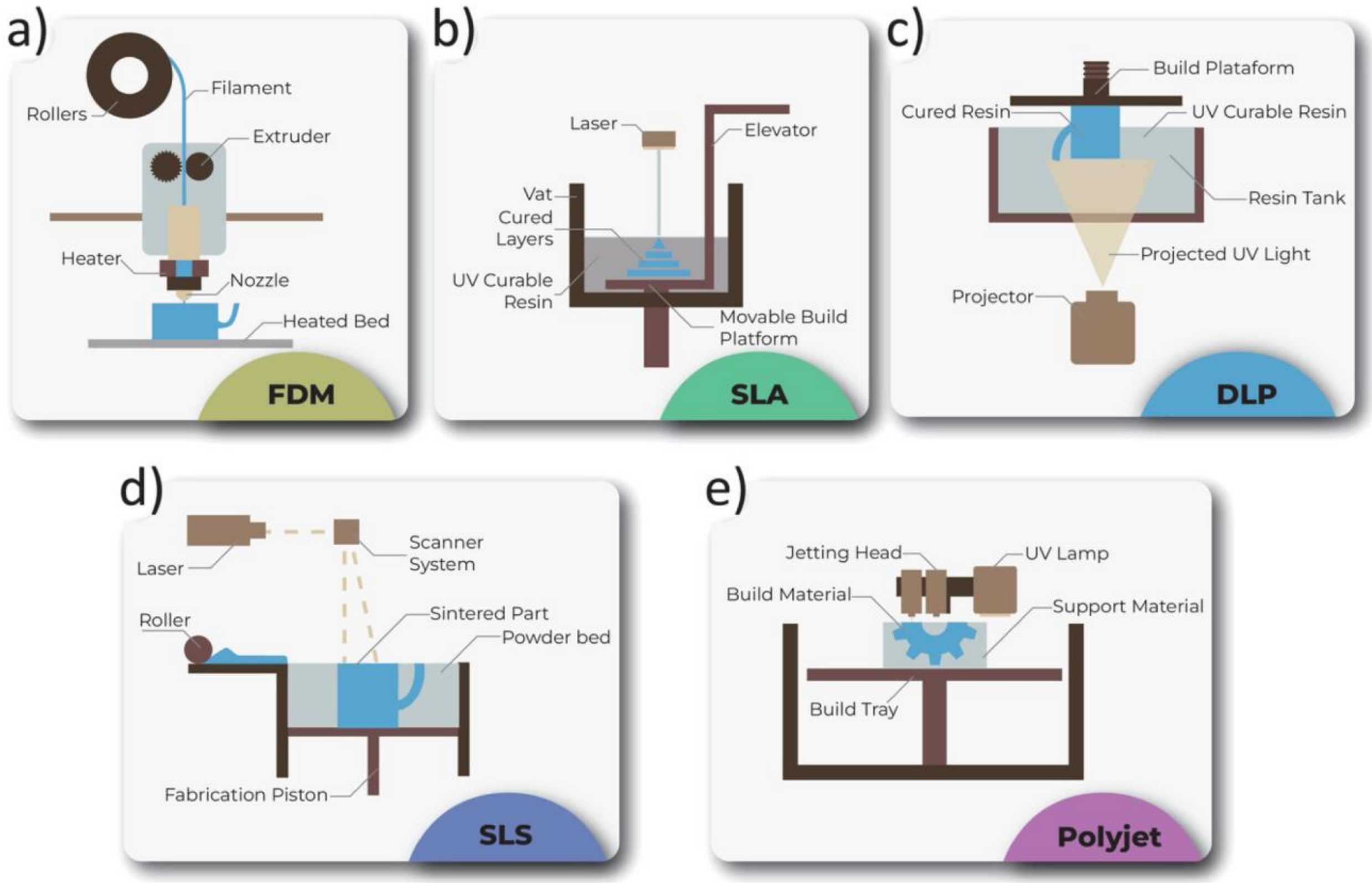

Figure 2. Illustrated scheme of: (a) FDM; (b) SLA; (c) DLP; (d) SLS; and (e) Polyjet. (Figures and Captions above are from reference[20] without change, under Creative Commons BY 4.0 License.[21])

**Table 1.** Comparison of 3D printing technologies. (This table is from reference[22] without change, under Creative Commons BY 4.0 License.[21])

| Method | Accuracy (mm) | Material Compatibility | Surface Finish | Print Speed | Strength | Limitation | Equipment Cost |
|---|---|---|---|---|---|---|---|
| FDM | ±0.05 –0.3 | Thermoplastic | Moderate | 0.04–0.15 m/s | Low cost, Robust | Layer lines | $200–$15,000 |
| SLA | ±0.025 –0.1 | Photopolymer | Excellent | 10–100 mm/h | High accuracy | Limited materials | $2500–$25,000 |
| DLP | ±0.025 –0.1 | Photopolymer | Excellent | 20–100 mm/h | Faster than SLA | Limited materials | $5000–$50,000 |
| SLS | ±0.05 –0.2 | Thermoplastics & composites | Rough | 10–50 mm3/s | Mechanical properties | Powder hazard | $30,000–$200,000 |
| Poly Jet | ±0.01 –0.085 | Photopolymer | Excellent | 0.1–0.2 m/s | Very High Resolution | Expensive | $50,000–$500,000 |
| DIW | ±0.1 –0.5 | Exceptionally wide | Moderate | 0.001–0.5 m/s | Print multi-materials | Viscosity requirement | $10,000–$100,000 |

In summary, AM has revolutionized the production and use of polymers, enabling higher performance and more complex geometries. While conventional polymer manufacturing has reached cost-performance limits, especially for very demanding thermo-mechanical properties, formative manufacturing remains the go-to for scale-up, with AM catching up for limited production. Many demanding applications can also face issues related to environmental sustainability and reduced plastic waste. AM should provide greater form freedom and rapid prototyping to improve time-to-market. [23–25]

## 2.1 From Industry 4.0 to Industry 5.0 and Digital Manufacturing

Many expectations for future industry sectors for the past 10 years has been inspired by the Industry 4.0 ecosystem.[26] Considering emerging technologies that will impact the history of industrialization, including AM and AI/ML, these principles have been transforming what are considered the latest technologies and protocols to improve efficiency and digitization. Industry 4.0 builds on contemporary and digital technologies, advanced computing, including flexibility and transferability, intelligent systems, and AI-driven and autonomous operations responsive to changing environments. Big data, blockchain technologies, and AI will enable the so-called Smart Factory, a digitally transformed industry. The next evolution of manufacturing, building on Industry 4.0's technologies (AI, IoT, robots) is now referred to as Industry 5.0.[26] It shifts the focus to human-centricity, sustainability, and resilience, emphasizing human-robot collaboration (cobots) for personalized production, ethical AI, reduced environmental impact, and resilient value chains, moving beyond pure automation to empower workers with creativity, not replace them. The aim is for a 'net positive' impact by aligning technology with societal well-being and ecological goals. Industry 4.0 and 5.0 will shape the future of manufacturing and new biobased materials, driving expansion and improving performance through technological advancements such as AI, digital twins, and ML. Thus, Industry 5.0 and Industry 4.0 technologies will advance together with digital technologies and new manufacturing regulations, where the human-in-the-loop will remain relevant. We envision the case for renewable bio-based materials and AM.

## 2.2 3D Printing and AM for a Circular Economy

Due to its digital and additive nature, 3D printing is a promising technology that offers wide-ranging opportunities to produce customized products, eliminate unsold inventory, and reduce overall inventory.[27] This on-demand or on-time production can save materials and other resources, enabling sustainable output with less waste. Not only does it enable on-demand fabrication, but it also facilitates on-site repair and replacement. This can be synergistic towards a circular economy, where 3D printing is part of the solution.[27] The concept of circular economy was proposed early for polymers.[28,29]

AM can adopt a more circular production approach by using biobased and renewable content, recycling, and upcycling materials as inputs for 3D printing processes.[30] Towards this goal, the following are needed: a) increased knowledge of 3D printing should be co-designed for circularity process efficiency; b) organizational shift in the supply chain of raw materials with localization and recycling operation; c) new genomes of materials classes with upcylability; c) opportunities for start-up companies with incentives; and d) training of skilled workforce and materials scientist in 3D printing and R&D of renewable and sustainable materials economy.

## 3.Bio-based materials and sustainability: 3D Printing and materials optimization

There is a strong interest in biobased materials for 3D printing, with a focus on sustainable, renewable resources.[31] This includes Polylactic acid (PLA), derived from corn or sugarcane; polysaccharides such as cellulose, starch, carrageenan, and chitosan; polyhydroxyalkanoates (PHAs); and other bio- and microbiologically derived polymers. These polymers have been used primarily for their classification as “natural polymers” and their availability. A comprehensive classification of various natural polymers, along with specific examples, is available in Figure 3. They offer a pathway for circularity, sustainability, biodegradability, and even biocompatibility and are abundant.

Many methods based on DIW, FDM, and other AM extrusion methods are the norm and require some reformulation to achieve proper viscoelastic and melt behavior.[32] Typically, the properties are not equivalent to formatively manufactured parts and require new chemical synthesis, melt blending, and incorporating filler materials.[33] The processing challenges differ from those of synthetic polymers. Typically, their low melting point and easy degradability or charring behavior can be a problem. Blending is another method to improve their 3D printability.[33] Sulapac (wood/biopolymers) is a bio-derived resin blend that is suitable for 3D printing.[34] Still, PLA is a popular material, and possibly PHA-based polyesters in the future.[32] PLA is derived from fermented plant starches (corn, sugarcane, palm) and can also be formulated with appropriate additives or a sufficient MW. PHA (Polyhydroxyalkanoates): A Biodegradable polymer from bacterial fermentation, offering better biodegradability than PLA.

Other polysaccharides, such as cellulose and starch-based polysaccharides, can be derived from wood pulp, plant fibers, and other agro-based raw materials and naturally offer renewability.[33] However, they require advanced processing and energy or industrial heat waste management. There is also interest in bio-based Nylons (e.g., PA11) for SLS 3D printing. Nylons are generally popular for their toughness, but they have a higher melting point and absorb moisture. Wood/Plant Composites[33] can be derived from biobased fibers or natural fibers blended with fossil-based synthetic polymers. The Filaments are essentially blends of wood flour or plant fibers, including Sulpac. Biobased resins, hydrogels, and photoreactive resins are also gaining popularity with SLA/DLP photopolymerization and bioprinting. These can also be in the form of nanocellulose or nanowhisker additives. Hydrogels (alginate, gelatin, guar, carrageenan, etc.) can also be used for materials extrusion printing, such as DIW, if they exhibit shear thinning, sufficient yield stress, and a high storage modulus.[32]

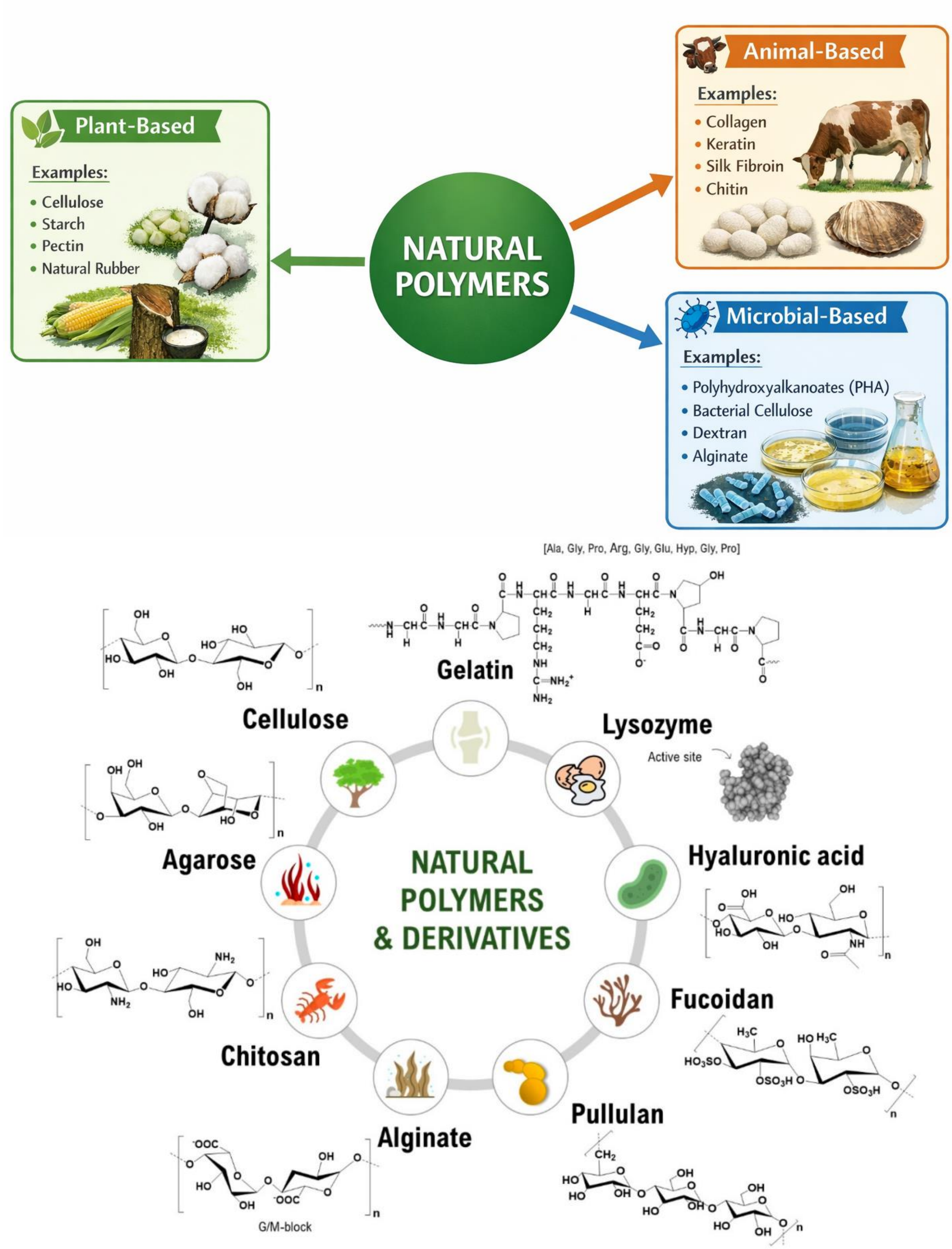

Figure 3. Top: Natural polymers by origin. Bottom: Examples of natural polymers and derivatives. (Figure and Caption are from reference,[35] without change, under a creative commons license.[21])

In more recent reviews,[36,37] we examined the broader classes of hydrogels, including biopolymers, synthetic polymers, and nanocomposites, and their future applications in new materials, 4D printing, and biomedical or bioengineering applications. The potential of polyelectrolyte complexes (PECs) and their coacervates has been demonstrated, offering the opportunity for more biobased polymers and hydrogel formulations for 3D printing. These PECs exhibit a range of humidity-dependent viscoelastic properties.[38]

Some examples of our previous work in bioderived polymers and nanocomposites included: 3D printing of blended thermoplastic polyurethane (TPU) with poly(lactic acid) (PLA) resulting in a thermo-mechanically optimized material. But with graphene oxide (GO), the nanocomposite becomes a superior material with biocompatible features.[39]

We successfully demonstrated the 3D printing of a nanocellulose nanocomposite hydrogel via a stereolithography apparatus (SLA), with high potential for biomedical engineering applications.[40] We have demonstrated improved properties in polymethacrylate composites using surfactant-modified chitosan (SMCS) particles at loadings between 2–10 wt%. Ionic complexation facilitated compatibility and dispersion by non-covalent bonds, resulting in improved thermo-mechanical properties.[41]

By utilizing chitin nano whiskers at as low as 0.5 wt%, we demonstrated superior properties in SLA 3D-printed methacrylate (MA) resins. We demonstrated enhanced thermomechanical and thermal properties in a 3D printed chitosan nanocomposite.[42]

The biodegradability of natural polymers and fibers can be an issue when replacing a synthetic polymer for the intended application.[43,44] Although ideal for new 3D printing materials, the sustainable and renewable sourcing advantages can also be offset by inferior thermomechanical properties or even higher cost. Typically, the cost burden is in the purification steps and the lack of quality control. Many natural fibers offer unique advantages for new 3D printing materials and composites. These fibers include hemp, bamboo, pineapple fibers, miscanthus grass, and others.[45–47] They serve as natural fillers or reinforcements that improve the tensile, compressive, modulus, and flexural properties, along with thermoplastics such as polystyrene (PS), polymethylmethacrylate (PMMA), PLA, acrylic-butadiene-styrene (ABS), and polypropylene (PP). Fiber-based composites can be used in many applications. However, various challenges with the bio-based fibers include a lack of scalable sources, moisture absorption, restricted processing temperatures, varying quality or polydispersity, and temperature instability. FDM and other extrusion-based processes can leverage this to achieve superior thermo-mechanical properties in fiber composites. The advantages of natural polymers and natural fiber blends are particularly promising for 3D printing, owing to their cost and synergistic mixing. There is extensive research on their application in many types of formative manufacturing, including injection molding, thermoforming, and coatings.[48,49] Polysaccharide-based or protein-based gels and viscoelastic materials can be fabricated by 3D printing for applications in biomedical engineering, tissue implants, and packaging. Cellulose-based natural fibers are common yet strong additives for composites.[50] Future research on these fibers for 3D printing (3DP) will continue to address this demand for high performance, and sustainability will remain a priority. Their intrinsic properties should be taken as an advantage: 1) Sustainability: made from renewable resources, reducing reliance on fossil fuels. 2) biodegradability: reduces plastic waste, 3) biocompatibility: ideal for medical/food-contact applications, and 4) reduced toxicity: PLA emits fewer harmful vapors than ABS during printing.

In summary, AM can be considered a benchmark for advancing the use and development of more bio-based materials in the plastics and composites industries for advanced manufacturing. The focus now is to go beyond lab- and pilot-scale research, with cost-effectiveness as an AM material. 3D printing as a manufacturing method, the challenges remain in poor interlayer adhesion, anisotropic mechanical properties, residual stress due to poor semi-crystallinity control, and poor finishing. This will require more AI/ML-driven workflows.

# 4.AI/ML Workflows and SDLs

With recent advances and applications in 3D printing and manufacturing, there is strong interest in incorporating more AI/ML-driven workflows across concept, design, and manufacturing.[51–53] AI/ML workflows will facilitate the development of new 3D-printing materials, protocols, and machines. Figure 4 provides an overview of supervised and unsupervised learning methods for biopolymer production. Figure 5 offers an overview of deep learning methods in biopolymer production. The in-situ, real-time use of sensors can provide a feedback loop for monitoring and control. In new polymers and composite synthesis, self-driving laboratories (SDL) will emerge as a high-throughput experimentation setup. Understanding the structure, composition, processing, and properties (SCPP) of new polymers and their function is critical. For 3D printing, optimization will lead to superior properties for demanding applications in aerospace, automotive, and armaments manufacturing, among others. The convergence of materials and AI/ML is inevitable. Table 2 compares ML and DL applications in polymer design, processing, and printing.

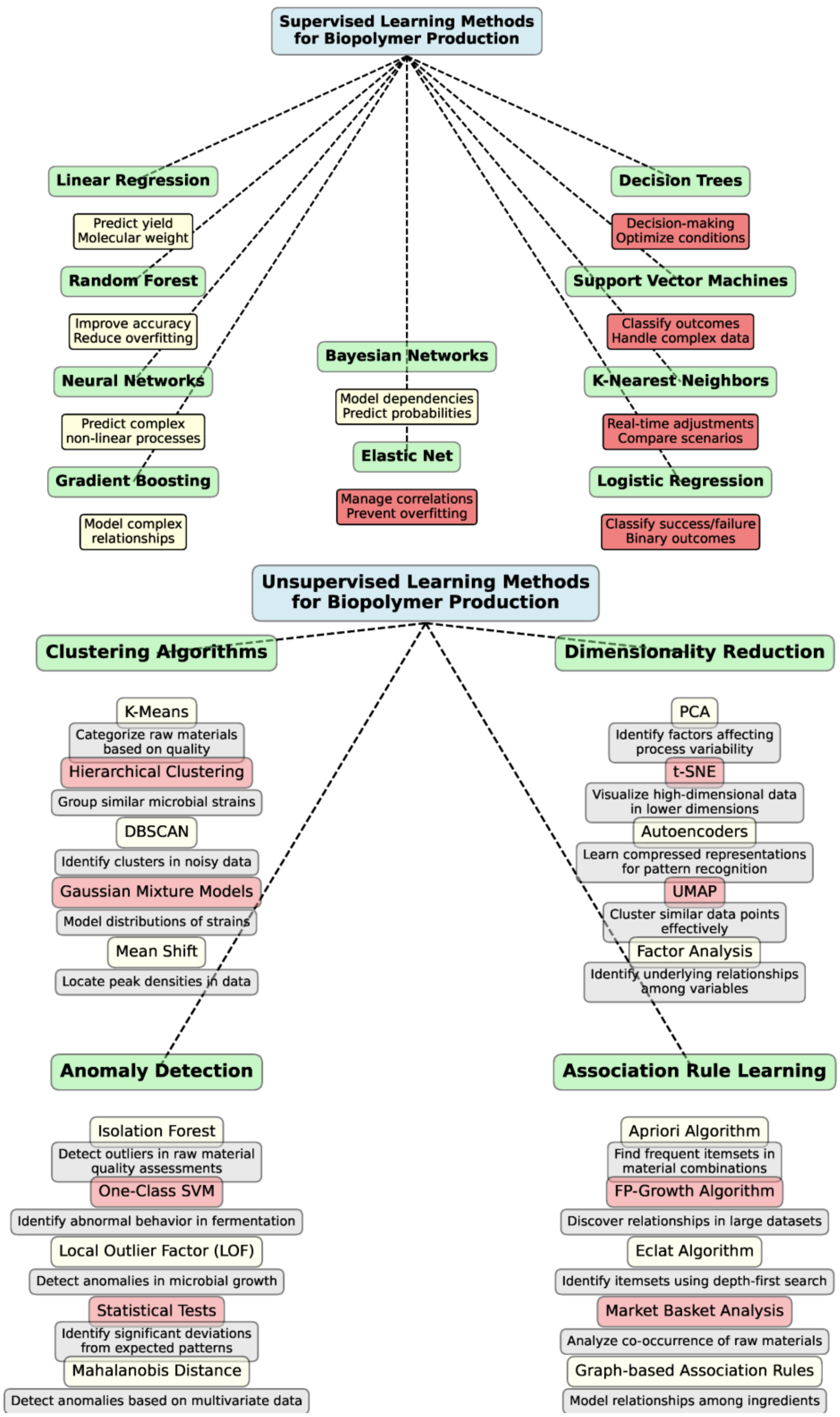

Figure 4. Top: Overview of supervised learning methods applied in biopolymer production. Bottom: Unsupervised learning methods in biopolymer research with possible applications. (Graphs and captions are from reference,[54] without change, under a creative commons license.[21])

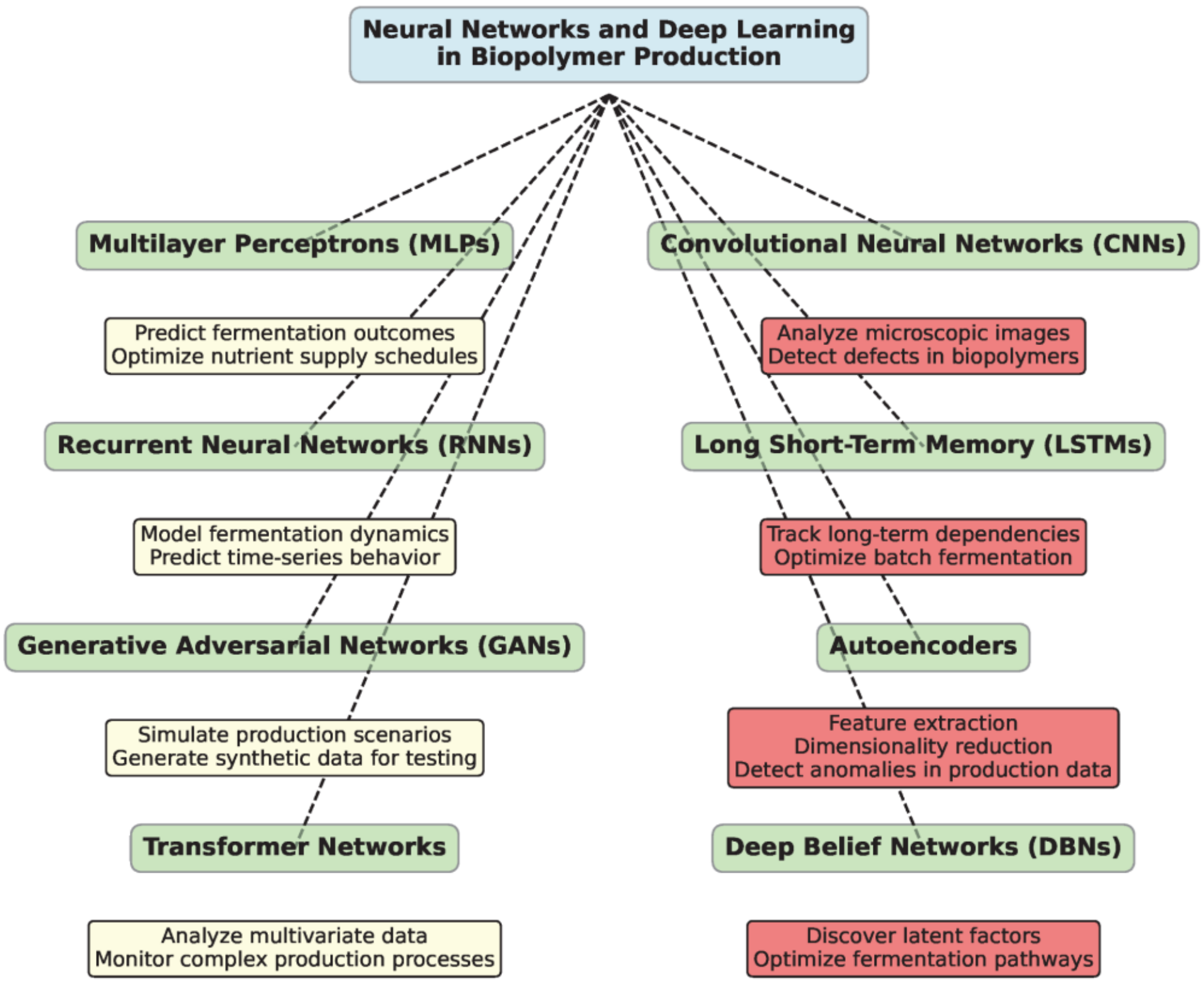

Figure 5. Application of various NNs and deep learning architectures in biopolymer production.(Graphs and captions are from reference,[54] without change, under a creative commons license.[21])

For new materials with AI/ML approaches, it is essential to design them at the outset for manufacturability. To achieve optimal polymer design (molecular weight -MW, melting point, branching architectures, Tg, etc.) for properties and processability, many experiments are typically required.[55] AI/ML can facilitate significant progress in polymer materials chemistry by combining polymer properties with the intended 3D printing method at the outset. By starting with a good Large Language Model (LLM) simulation, it is possible to augment the concept of Self-Driving Labs (SDL) in AI/ML workflows.[56]

Table 2. Comparing ML and DL applications in Polymer design, Processing and Printing[52,57–62]

| | **Polymer design** | **Polymer processing** | **3D/4D printing** |
|---|---|---|---|
| Objective | Discover new polymers with target properties. | Optimize continuous extrusion, molding. | Achieve accurate, defect- free printed parts and smart responses. |
| Bottleneck | Small, heterogeneous property datasets; synthetic data often used. | Need labeled quality data and robust sensor integration. | Coupling multi- modal data (materials, G- code, images, time series). |
| Inputs | Molecular structure, composition, process conditions. | Machine settings, sensor time series, material grades. | Material formulation, print parameters, geometry, in- situ images/signals. |
| Labels | $T_g$, modulus, conductivity, bandgap, biocompatibility. | Dimensional accuracy, surface quality, throughput, defect rates. | Printability scores, defects, porosity, mechanical/biological performance. |
| ML tools | Random forest (RF) / Gradient Boost, Graphic NNs, transformers, inverse- design frameworks. | RF/Neural Networks (NN) regressors, fault- diagnosis ML, physics- informed NNs. | Artificial NNs for printability, CNNs/vision transformers, Bayesian optimization, RL. |

We have demonstrated ML methods in sensor materials and nanocomposites.[63,64] This was done by correlating DL with the experimental design and the composite's composition. We have also used ML methods, including simulation, to look at the optimal composition and 3D printing of piezoelectric nanocomposites.

Significant advances with AI/ML workflows in materials research are leaning towards the 'self-driving laboratory' (SDL).[65,66] The SDL enables more transformative use of autonomous laboratory tools, including sensors, edge computers, robotic and programmable handling, API developments, and, in general, more agentic AI tasks. While not readily devoid of a human operator, an SDL can be demonstrated in the laboratory through synthesis and characterization. With synthesis and continuous-flow chemistry reactors, it is possible to perform more ML-driven reaction engineering. [67,68] In a recent paper, ML was combined with an automated continuous-flow chemistry synthesis protocol for polylactide (PLA) via an organocatalyzed reaction mechanism.[69]

A significant development in new biobased materials for 3D printing can be achieved through continuous-flow reaction chemistry and polymerization. New types of hypothesis-driven goals can include: 1) molecular weight distribution (MWD) and MW, 2) copolymers and graft architectures, 3) hyperbranching, 4) chirality, 5) functional group density with polymer analogous reactions, and 6) in an SDL setup, chemical engineering unit operations can be performed to optimize kinetic and reaction parameters, including composition volume (V), flow rate, pressure (P), temperature (T), and reagent equivalents, in space or time series. An important, immediate goal is high-yielding reaction parameters. This has been demonstrated for acrylate-vinyl copolymerization chemistries.[70]

With mechatronics and computer numerical control (CNC) motion, an automated SDL setup is possible. In the flow reactor, the autonomous design will enable the simultaneous exploration of different reagents, solvents, monomers, and reaction conditions (P, V, and T) in each experiment. It is possible to exploit online characterization to give real-time or continuous feedback-loop protocols.[71] For the flow reactor, the autonomous design will enable different reagents, catalysts, additives, or reaction parameters for each experiment (P, V, and T). Using a flow reactor also results in safer experiments and demonstrates scale-up starting at the bench scale.[72]

The reaction, which can be performed under homogeneous or heterogeneous conditions (packed-bed or plug-flow column), can be monitored in real time using several analytical spectroscopic and chromatographic instruments. Each can serve as part of a feedback loop to adjust the model and training instructions for the data. In an SDL set-up, massive amounts of data are inevitable. The monitoring sensors can include infra-red (IR), Raman, nuclear magnetic resonance (NMR), electrochemical, etc.[73,74] This can be used to provide "cleaner" original data, available for further LLM build-up and instructions. With the implementation of a continuous feedback loop, the SDL setup becomes a real-time, even autonomous, system for data collection and training. Interfacing with higher computer resources, such as the cloud for both storage and fast computing, will be necessary. This SDL setup will support researchers in synthesizing, analyzing, and interpreting field data for biobased polymer materials and 3D printing.[75,76]

Some examples of advances in the use of continuous-flow reactors, AI/ML- driven workflows, and SDL-optimized synthesis of biobased and biodegradable polymers have been previously reported.[71,77–83]

SDL can also be applied to 3D printing. While ML can optimize materials and designs for 3D printing, it can also be used in real time with an SDL. This includes real-time methods for removing residual stress, defect detection, such as air, self-correcting viscosity changes with flow rates, and adjusting heating or photocuring parameters. With a feedback loop to correct defects during 3D printing, an optimal process is realized with each printing, resulting in optimized thermo-mechanical properties. The challenge is to design this SDL with sensors and computational resources and to include instructions ( mechatronics) to enable autonomous operation. Again, access to cloud storage and high-performance computing (HPC) will result in data generation that can be used for future LLMs

Several examples of ML-optimized autonomous 3D printing operations can be found in the literature.[84–87] More AI/ML optimizations for biobased polymer printing have yet to be demonstrated, but the number of works on 3D printing optimization is expected to grow.[88]

# 5. Outlook and Conclusion

In summary, AI/ML will enable new polymer synthesis and 3D printing of polymers at a higher level of research for both discovery science and optimization. The use of new tools and even skill sets based on ML-driven design and process intervention will lead to more bio-based polymers and filler materials being used for AM. The following are immediate directions that can be pursued: 1) build databases and LLMs on current research and data of bio-based polymers and composites with the goal of targeting key hypothesis-driven objectives, 2) co-design new 3D printing methods to accommodate the advantages of using bio-derived materials, 3) build more autonomic and physical AI tools, including SDL-based automation and workflows, to scale up laboratory research, and 4) employ generative AI and DL to look at new types of materials and even bio-inspired design for emerging applications in sensors, biomedical engineering, and extreme environment materials (nuclear, space, mining, etc.). This should lead to more discovery science, higher-throughput experimentation, faster process optimization, and more economical integration towards Industry 4.0 and 5.0 adaptation with bio-based materials.

## Acknowledgement

This work was supported by the Center for Nanophase Materials Sciences (CNMS), which is a US Department of Energy, Office of Science User Facility at Oak Ridge National Laboratory, and Laboratory Directed R&D (ORNL INTERSECT).

## Conflicts of Interest and Declarations

The authors declare no conflict of interest. Also, an archived version of the manuscript was submitted as a non-peer-reviewed preprint in an archiving portal.

## Author Contributions

RCA planned the layout, and drafted Section 1-5. JC contributed to Section 4-5, and the overall design and writing.

Last update: 2.18.2026